\documentclass[amsmath,amssymb,prc,twocolumn,english,showpacs]{revtex4}
\usepackage[T1]{fontenc}
\usepackage[latin1]{inputenc}
\usepackage{babel}
\usepackage{graphics}
\usepackage{comment}

\makeatletter

\usepackage[T1]{fontenc}
\usepackage[latin1]{inputenc}
\usepackage{babel}
\usepackage{graphics}

\newcommand{\ra}{\rangle}
\newcommand{\la}{\langle}
\newcommand{\rp}{\right)}
\newcommand{\lp}{\left(}
\newcommand{\rc}{\right]}
\newcommand{\lc}{\left[}

\newcommand{\C}{{}^{12}\mathrm{C}}



\makeatother
\begin{document}

\title{Two-body scattering without angular-momentum decomposition}

\author{M. Rodr\'{\i}guez-Gallardo$^{1}$, A. Deltuva$^{1}$, E. Cravo$^{1}$, R. Crespo$^{1,2}$, and A.~C. Fonseca$^{1}$}

\affiliation {$^1$ Centro de F\'{\i}sica Nuclear, Universidade de Lisboa, Av. Prof. Gama Pinto 2, 1649-003 Lisboa, Portugal}
\affiliation {$^2$ Departamento de
F\'{\i}sica, Instituto Superior T\'ecnico, Taguspark,\\
Av.\ Prof.\ Cavaco  Silva,  
2780-990 Porto Salvo, Oeiras, Portugal}

\date{\today}

\begin{abstract}
Two-body scattering is studied by solving the Lippmann-Schwinger equation
in momentum space without angular-momentum decomposition
for a local spin dependent short range interaction plus 
Coulomb. 
The screening and renormalization approach is employed 
to treat the Coulomb interaction.
Benchmark calculations are performed by comparing our procedure
with partial-wave calculations in configuration space for
$p$-${\rm^{10}Be}$, $p$-${\rm^{16}O}$ and $\C$-${\rm^{10}Be}$
elastic scattering, using a simple optical potential model.
\end{abstract}
\maketitle

\section{Introduction}

The aim of the present work is to solve the two-body Lippmann-Schwinger equation without partial-wave decomposition for a local short range interaction plus Coulomb.
This is a first step towards the ultimate goal of solving exact three-body equations 
without partial-wave decomposition as a means to describe complex nuclear reactions where three-body degrees of freedom play a significant role.

The inclusion of the long range Coulomb force between charged particles of equal sign has become possible, in recent years, through a novel implementation of the 
method of Coulomb screening and renormalization~\cite{Tay75,Alt78} in the framework of Alt, Grassberger and Sandhas (AGS) exact three-~\cite{Alt67} and four-body~\cite{grassberger:67} integral equations, leading to fully converged results for three-\cite{De05a,Del05d} and four-nucleon scattering~\cite{Del07a,Del07b} and for direct nuclear reactions dominated by three-body degrees of freedom~\cite{deltuva:06b,Cr07,De07,Crespo07}.

In all these calculations~\cite{De05a,Del05d,deltuva:06b,Del07a,Del07b,De07,Cr07,Crespo07} the equations were solved using the partial-wave decomposition of the multivariable integral equations and taking as many partial waves as needed for convergence of the observables. Although we get fully converged results, at intermediate energies the partial wave expansion converges very slowly and may get unstable if we increase the energy beyond the values we used or if we address reactions with two heavier nuclei such as $\C$-${\rm^{11}Be}$ where $\C$ and ${\rm^{10}Be}$ are considered as inert cores. In this case, even at moderate energies the relative $\C$-${\rm^{11}Be}$ wave length is so small that a very large number of partial waves is required for convergence. 

Therefore, in order to get exact three-body results at higher energies or involving the collision of two heavy nuclei one may need to develop calculations without partial-wave decomposition. The group at Ohio University has already made progress in this direction for two-body ~\cite{El98,Fa00} and three-body scattering
~\cite{Sc00,Liu05}, following an earlier work by Belyaev et al.~\cite{Bel75}. However they do not include the exact treatment of Coulomb in their calculations, 
though in their early works~\cite{Ch91,El93,Ch95} they introduced Coulomb in an approximate way
in the context of a  multiple scattering framework.

In the present manuscript, we show results for the solution of the two-body Lippmann-Schwinger equation for $p$-${\rm^{10}Be}$, $p$-${\rm^{16}O}$ and $\C$-${\rm^{10}Be}$  elastic scattering at intermediate energies. The novelty of the present work vis-a-vis the Ohio group two-body calculations is that we include Coulomb between two charged cores of mass $A_i$ and atomic number $Z_i$ together with standard optical potentials that include a central plus a spin-orbit interaction.

This paper is structured as following. In Sec.~II we present the formalism of the Lippmann-Schwinger equation for central and the spin-orbit potentials. In Sec.~III we summarize the Coulomb treatment. In Sec.~IV we show the results for different reactions. In Sec.~V we summarize and give conclusions. Finally in Appendix~A we present  the analytical expressions for the Fourier transform of the potentials used in this work and in Appendix~B we explain the method used to solve the Lippmann-Schwinger equation. 

\section{Lippmann-Schwinger equation}

The Lippmann-Schwinger equation for the scattering of two particles is 
\begin{equation}
T=V+VG_0T,
\end{equation}
where $V$ is the two-body potential between the particles, $G_0=(Z-H_0)^{-1}$ the free two-body resolvent, and $T$ the transition operator. The matrix element of the transition amplitude in momentum space, $T({\bf q}',{\bf q},Z)\equiv\la{\bf q}'|T(Z)|{\bf q}\ra$, satisfies the integral equation
\begin{eqnarray}
  \lefteqn{T({\bf q}',{\bf q},Z)=V({\bf q}',{\bf q})}\label{lse} \\ \nonumber
  &+&\int d^3q''V({\bf q}',{\bf q}'')\frac{1}{Z-\frac{\hbar^2q''^2}{2\mu}}T({\bf q}'',{\bf q},Z).
  \end{eqnarray}
Here, ${\bf q}$ is the relative wave vector, $\mu$ is the reduced mass of the two particles and Z 
the appropriate energy. 

\subsection{Central interaction}

First we consider the case of a central interaction only where  
we follow the same procedure as Elster et al.~\cite{El98}.
In this case, the matrix elements in momentum space of the potential and the transition amplitude, 
$V({\bf q}',{\bf q})$ and $T({\bf q}',{\bf q},Z)$ respectively, are scalar functions
\begin{eqnarray}
V({\bf q}',{\bf q})&=&V(q',q,{\bf \hat{q}}'\cdot{\bf\hat{q}}),\\
T({\bf q}',{\bf q},Z)&=&T(q',q,{\bf \hat{q}}'\cdot{\bf \hat{q}},Z).
\end {eqnarray}
Therefore Eq.~(\ref{lse}) can be expressed as follows
\begin{eqnarray}
\lefteqn{T(q',q,x',Z)=V(q',q,x')\label{lse-2}+\int_0^{\infty} dq''q''^2\int_{-1}^1dx''} \nonumber \\
&\times&\int_0^{2\pi}d\varphi'' V(q',q'',y)\frac{1}{Z-\frac{\hbar^2q''^2}{2\mu}}T(q'',q,x'',Z),
\end{eqnarray}
where $x'={\hat {\bf q}}'\cdot{\hat {\bf q}}$, $x''={\hat {\bf q}}''\cdot{\hat {\bf q}}$, 
and $y={\hat {\bf q}}''\cdot{\hat {\bf q}}'$. 
We take the incoming wave vector ${\bf q}$ in the direction of the z-axis and the arbitrary azimuthal angle $\varphi'$ for ${\bf q'}$ is chosen to be zero. We can then express $y$ through $x'$ and $x''$ as
\begin{equation}
y=x'x''+\sqrt{1-x'^2}\sqrt{1-x''^2}\cos{\varphi''}.
\label{yx}
\end{equation}
Defining
\begin{equation}
v(q',q'',x',x'')\equiv \int_0^{2\pi}d\varphi'' V(q',q'',y),
\label{ipot}
\end{equation}
the integral Eq.~ (\ref{lse-2}) becomes
\begin{eqnarray} \label{eq:T}
\lefteqn{T(q',q,x',Z)=\frac{1}{2\pi}v(q',q,x',1)+\int_0^{\infty} dq''q''^2}\label{lse-3}\\ \nonumber
&\times&\int_{-1}^1dx''v(q',q'',x',x'')\frac{1}{Z-\frac{\hbar^2q''^2}{2\mu}}T(q'',q,x'',Z),
\end{eqnarray}
leading to a  two-dimensional integral equation in the off-shell wave vector $q''$ and the cosine of the scattering angle $x''$.

\subsection{Spin-orbit interaction}

Next, we consider the case in which 
we have a spin-orbit interaction. In this case, Eq. (\ref{lse}) becomes a set of coupled equations
\begin{eqnarray} \label{eq:Ts'}
\lefteqn{T_{s'\lambda',s\lambda}({\bf q}',{\bf q},Z)=V_{s'\lambda',s\lambda}({\bf q}',{\bf q})}\label{lses}\\\nonumber
&+&\sum_{s''\lambda''}\int d^3q''V_{s'\lambda',s''\lambda''}({\bf q}',{\bf q}'')\\ \nonumber
&\times&\frac{1}{Z-\frac{\hbar^2q''^2}{2\mu}}T_{s''\lambda'',s\lambda}({\bf q}'',{\bf q},Z),
\end{eqnarray}
where $s$ is the spin of the system and $\lambda$ its projection in the z-axis.
The spin-orbit potential,
commonly expressed  as  
\begin{equation}
V_{so}({\bf r})=v_{so}(r)~{\bf \sigma}\cdot{\bf l},
\label{sop}
\end{equation}
is not central anymore. The ${\bf \sigma}\cdot{\bf l}$ term introduces a dependence on the azimuthal 
angle $\varphi'$ which makes Eq.~(\ref{eq:T}) not valid. 
Nevertheless, it is possible to reduce Eq.~(\ref{eq:Ts'}) to a two-variable integral equation.
From the partial-wave analysis of the T-matrix
\begin{eqnarray}
\lefteqn{T_{s'\lambda',s\lambda}({\bf q}',{\bf q},Z)= \sum_{J M_J}\sum_{ L' M'}\sum_{ L M}
Y_{L'M'}(\hat{\bf q}')}\\ \nonumber
&\times& \langle L' M' s'\lambda' | J M_J \rangle T_{J}^{L's'Ls}(q',q,Z)
\langle L M s\lambda | J M_J \rangle Y_{LM}^{\ast}(\hat{\bf q}), 
\end{eqnarray}
assuming, as before, the initial wave vector ${\bf q}$ to be along the z-axis, it follows that
the $\varphi'$-dependence of $T_{s'\lambda',s\lambda}({\bf q}',{\bf q},Z)$ is determined by the spherical harmonics $Y_{L'M'}(\hat{\bf q}')$
in terms of  $e^{iM'\varphi'}$ with fixed $M'=\lambda-\lambda'$ because $M=0$.
Therefore the T-matrix can be written in factorized form as
\begin{equation} \label{t-phi}
T_{s'\lambda',s\lambda}({\bf q}',{\bf q},Z) = e^{i(\lambda-\lambda')\varphi'} 
\mathcal{T}_{s'\lambda',s\lambda}(q',q,x',Z)
\end{equation}
where $\mathcal{T}_{s'\lambda',s\lambda}(q',q,x',z)$ is the solution of a set of two-variable integral equations
\begin{eqnarray}
\lefteqn{\mathcal{T}_{s'\lambda',s\lambda}(q',q,x',Z)=\frac{1}{2\pi}v^{\lambda}_{s'\lambda',s\lambda}(q',q,x',1)}\label{lses-3}\\ \nonumber
&+&\sum_{s''\lambda''}\int_0^{\infty} dq''q''^2\int_{-1}^1dx''v^{\lambda}_{s'\lambda',s''\lambda''}(q',q'',x',x'')\nonumber\\ \nonumber
&\times&\frac{1}{Z-\frac{\hbar^2q''^2}{2\mu}}\mathcal{T}_{s''\lambda'',s\lambda}(q'',q,x'',Z).
\end{eqnarray}
Here $v^{\lambda}_{s'\lambda',s''\lambda''}(q',q'',x',x'')$ includes the phase from the T-matrix and $\varphi'$ is chosen to be zero
\begin{eqnarray}
\lefteqn{v^{\lambda}_{s'\lambda',s''\lambda''}(q',q'',x',x'')}\label{ipots}\\ \nonumber
&\equiv&\int_0^{2\pi}d\varphi''~e^{i(\lambda-\lambda'')\varphi''}V_{s'\lambda',s''\lambda''}({\bf q}',{\bf q}'')|_{\varphi'=0}.
\end{eqnarray}

As discussed in Ref.~\cite{El98,El99,Sc00,Fa00} these calculations are time consuming if the potential is given in configuration space and the transform to momentum space (Fourier transform) is performed numerically. Therefore in the present work we develop in Appendix~\ref{ApendA} analytic  Fourier transforms 
for Woods-Saxon interactions, central, surface, and spin-orbit, together with the screened Coulomb interaction.
In Appendix~\ref{ApendB} we outline the numerical procedure we follow to solve the integral equations.

\section{Treatment of Coulomb interaction}

The inclusion of the Coulomb interaction in momentum space is 
a very complicated task due to its $1/q^2$ singularity which together with the $G_0(Z)$ singularity renders the kernel of the Lippmann-Schwinger equation non-compact.
This fact has been a handicap for performing momentum space scattering calculations 
involving charged particles.
Over the years several methods  have been proposed to overcome
the Coulomb singularity
that  introduce a cutoff parameter: the pioneer work was done by
Vincent and Phatak~\cite{Vi74} which is, in principle, exact and works well for proton-proton scattering but does not yield sufficiently precise results for proton-nucleus scattering at intermediate energies~\cite{Cr90}  where accuracy in the high partial waves is needed for convergence; in addition this method cannot be extended to the three-body problem.
An improved method was also developed in Ref.~\cite{Cr90} which is capable
of producing more accurate quantitative calculations  than Vincent and Phatak for the reaction observables
for smaller values of the cutoff radius but still converges slowly with the cutoff radius for large scattering angles. 
An alternative method was proposed in Ref.~\cite{Ch91}
in which the limits of the cutoff parameter are taken analytically. More recently a novel technique was proposed~\cite{Oryu06} but its application is still limited to the numerical solution of the pure Coulomb problem~\cite{Oryu07}, and its numerical accuracy for high partial waves is yet to be tested.

\begin{figure}
\resizebox{0.45\textwidth}{!}{%
  \includegraphics{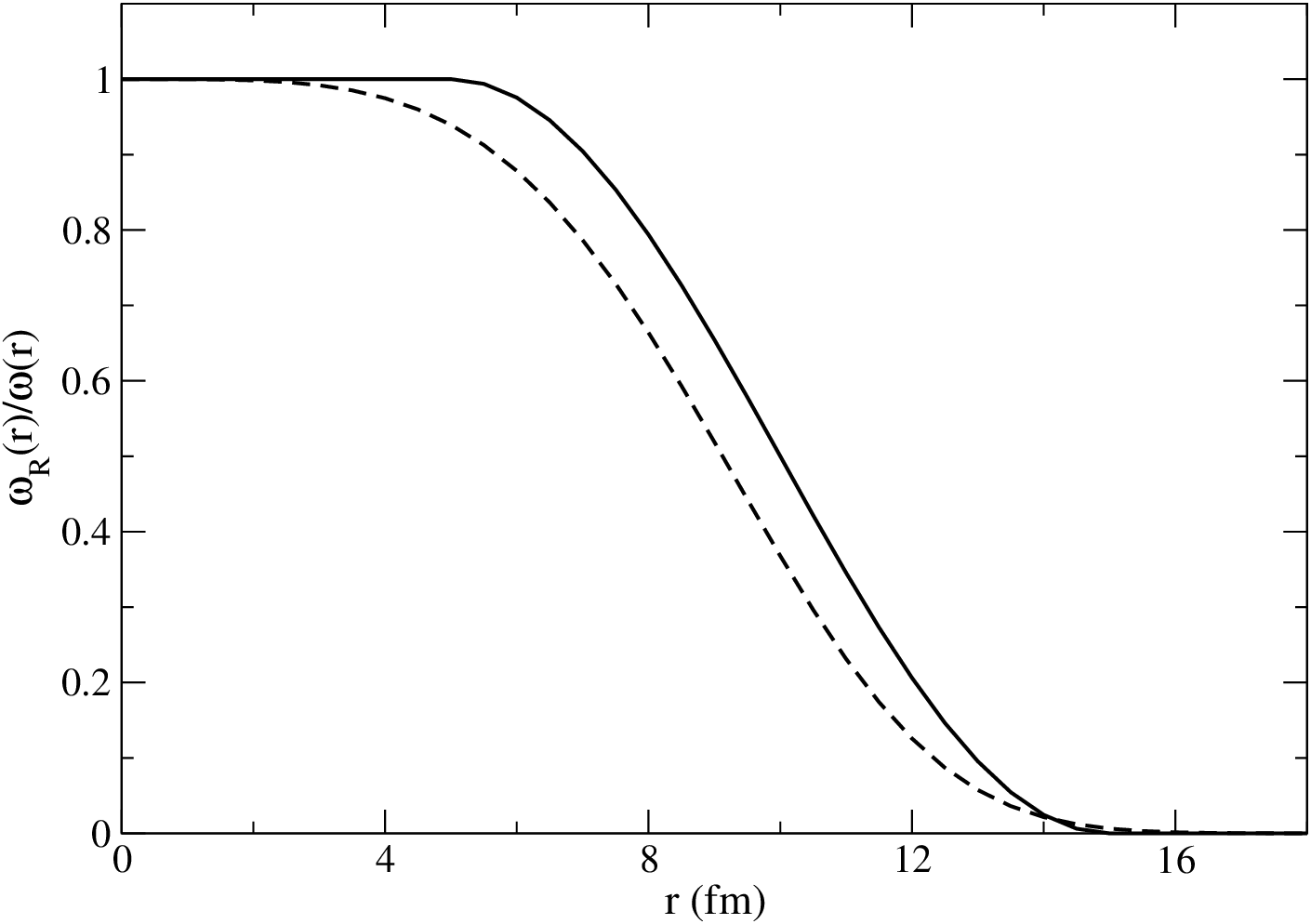}
}
\caption{Comparison between the screening function used in this work with $R=5$ fm (full line) and the screening function used in Ref.~\cite{De05a} with $R=10$ fm and $n=4$ (dashed line).}
\label{sf-comp}
\end{figure}

The screening and renormalization approach~\cite{Tay75,Alt78} has also been recently revisited leading to the treatment of the Coulomb interaction proposed in Ref.~\cite{De05a} together with a new screening function. For completeness we present in here a summary of the procedure. First, we work with a Coulomb potential $\omega_R$, screened around the separation $r=R$ between the two charged particles. In this work, we choose a screening function that is different from the one used in Refs.~\cite{De05a,Del05d,deltuva:06b,Del07a,Del07b}
\begin{subequations}
\begin{align} \nonumber
\omega_R(r) = {}& \omega(r)\lc\Theta(R-r) \right.  \\ 
 + {}& \left. \frac{1}{2}\Theta(r-R)\Theta(3R-r)\lp 1+\sin{\lp\frac{\pi r}{2R}\rp} \rp \rc , \label{scpot} \\ 
\omega(r) = {}& \frac{\alpha_eZ_pZ_t}{r} \; , 
\end{align}
 \end{subequations}
where $\alpha_e$ is the fine structure constant and $Z_p$ and $Z_t$ the ratio to the proton charge for both projectile and target nuclei. Unlike the screening function used before~\cite{De05a,Del05d,deltuva:06b,Del07a,Del07b}, this 
one has an analytical Fourier transform. In addition it possesses the same properties as the previous one, i.e., preserves the Coulomb interaction at short distances and for $r>R$ goes smoothly to zero. 
In Fig.~\ref{sf-comp} the shape of this new screening function
is compared  with the previous one, $\lp\omega_R(r) = \omega(r) e^{-(r/R)^n}\rp$, for $n=4$. 
The screening radius for Eq.~(\ref{scpot}) corresponds approximately to the double of the former one.
The screening radius $R$ is chosen to be larger than the range of the strong interaction.
However, it will be always very small compared with the nuclear screening distances which are of atomic scale (i.e., $10^5$ fm). Thus, the employed screened Coulomb potential $\omega_R$ is unable to simulate the physics of nuclear screening or even model all features of the true Coulomb potential. 
However following the prescription given in 
Ref. \cite{Tay75} and the technical developments proposed in Ref.~\cite{De05a}, the results corresponding to unscreened Coulomb can be obtained. 
This procedure involves the use of a two-potential formula that separates the long range part from the Coulomb modified short range contribution. Therefore the  amplitude $T$ for nuclear plus Coulomb scattering reads
\begin{equation}\label{eq:T=Tc}
T=T_c+\lim_{R\to\infty}\lc z_R^{-1/2}\lp T^{(R)}-T_c^{(R)} \rp z_R^{-1/2}\rc,
\end {equation}
where $T_c$ is the pure Coulomb amplitude that is 
known analytically and is given in the Appendix \ref{ApendA1}.
The pure Coulomb transition matrix that has no on-shell limit is not needed 
in the method of screening and renormalization \cite{Tay75,Alt78,De05a}.
The amplitudes $T^{(R)}$ and $T_c^{(R)}$ are calculated with
the nuclear plus screened Coulomb potential and the screened Coulomb potential
alone, respectively, as described in Sec.~II.
The second term on the right side of Eq.~(\ref{eq:T=Tc}) corresponds to the Coulomb modified nuclear short range amplitude which is calculated numerically for different $R$, but whose $R \to \infty$ limit is reached with high accuracy at finite $R$~\cite{De05a}. The renormalization factor $z_R$ in Eq.~(\ref{eq:T=Tc}) is given by
\begin{eqnarray}
z_R&=&\exp{\lp-2i\phi_R\rp},
\end{eqnarray}
where  $\phi_R$  is defined in Appendix \ref{ApendA1}. 

Unlike the work in Ref.~\cite{Oryu06, Oryu07} the method of screening and renormalization has limitations at low energy because the screening radius has to be at least greater than the wavelength associated with the relative motion  of two particles in the initial state. The lower the energy the higher the screening radius needed for convergence leading to numerical instabilities. Nevertheless for the typical energies of direct nuclear reactions the method is extremely accurate as demonstrated  in the following section where convergence is reached at finite screening radius ($R \le 12$ fm). Although there is no a priori way to predict the screening radius that leads to a converged result, we find that convergence is readily obtained by monotonically increasing $R$ until the calculated observables change by less than a given percentage value, typically $1\% $ or less.

\section{Results}

In this Section we are going to apply the formalism presented above to different two-body reactions. 
In order to show that the method works properly for a wide range of energies and observables
we consider first the calculation of $p$-${\rm^{10}Be}$ and $p$-${\rm^{16}O}$ elastic scattering observables obtained with a 
projectile-target optical model potential plus Coulomb at different  energies 
($E_{lab}=$50, 100, 150 and 200 MeV).
The results are compared with the solutions obtained by solving the Shr\"odinger equation in configuration space, where the screening of the Coulomb potential is not needed.
These calculations are performed with the code  {\sc fresco}~\cite{Thom88}
 which uses the standard method described in most Quantum Mechanics text books where, in each partial wave, the numerical solution of the differential equation for nuclear plus Coulomb potentials is matched, at some radius, with the appropriate asymptotic Coulomb wave functions which are known analytically. Convergence of results has to be tested with respect to the number of partial waves included and the matching radius . The numerical accuracy of this method is well known and is documented in Ref.~\cite{Thom88}.

For both reactions, we have used the parametrized optical potential of Watson~\cite{Wat69} at the corresponding energy per nucleon. The screening radius is taken to be 
$R=5$ fm for $p$-${\rm^{10}Be}$, and $R=7$ fm for $p$-${\rm^{16}O}$. In order to perform the numerical integrations we need to introduce a certain number of mesh points as explained
in Appendix~\ref{ApendB}. For the $p$-${\rm^{10}Be}$ reaction we take
$n_{\varphi}=$40, $n_{\theta}$=64, and $n_q$=64 and for $p$-${\rm^{16}O}$ we use 
$n_{\varphi}=$50, $n_{\theta}$=80, and $n_q$=80.
 The CPU time needed for convergence on a AMD Opteron (2.4 GHz) single processor  is about 10 minutes for $p$-${\rm^{10}Be}$ and 30 minutes for $p$-${\rm^{16}O}$. 

\begin{figure}
\resizebox{0.45\textwidth}{!}{%
  \includegraphics{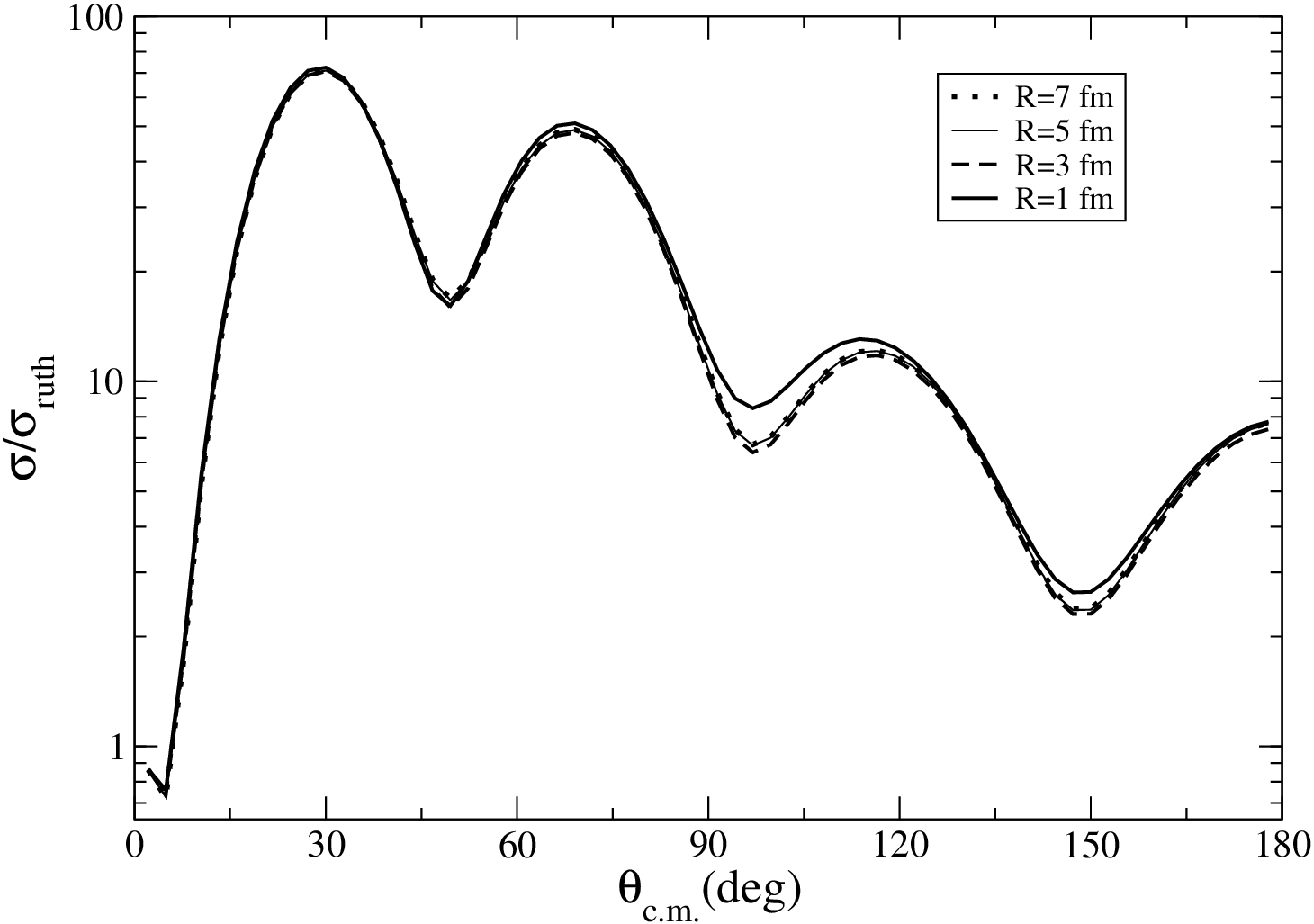}
}
\caption{ Differential cross section relative to Rutherford cross section for $p$-${\rm^{10}Be}$ elastic scattering at 50 MeV, calculated with different values of the screening radius $R$.}
\label{p10Be_conR}
\end{figure}

First, we show the convergence of the method itself with respect to the screening radius $R$ in Fig.~\ref{p10Be_conR}. The different lines are the differential elastic cross section relative to the Rutherford cross section for the reaction $p$-${\rm^{10}Be}$ at 50 MeV as the screening radius increases. From $R=5$ fm all curves fall on top of each other showing the convergence with $R$ and the numerical stability of the calculation.

\begin{figure}
\resizebox{0.45\textwidth}{!}{%
  \includegraphics{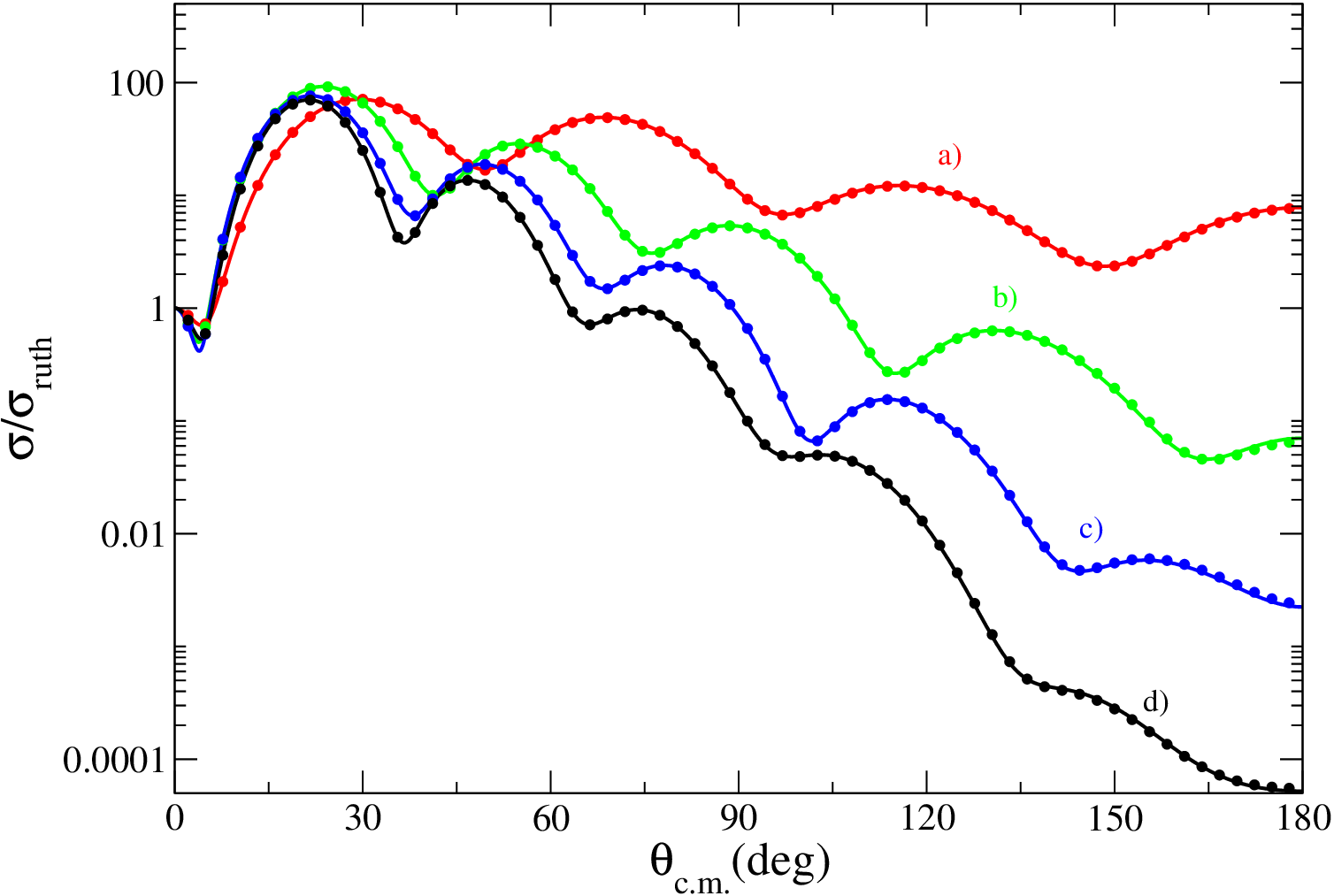}
}
\caption{(Color online) Differential cross section relative to Rutherford cross section for $p$-${\rm^{10}Be}$ elastic scattering. Full lines are partial-wave calculations in configuration space and points are plane-wave calculations in momentum space. The lines a), b), c), and d) correspond to $E_{lab}=$50, 100, 150, and 200 MeV, respectively.}
\label{p10Be_elR}
\end{figure}

\begin{figure}
\resizebox{0.45\textwidth}{!}{%
  \includegraphics{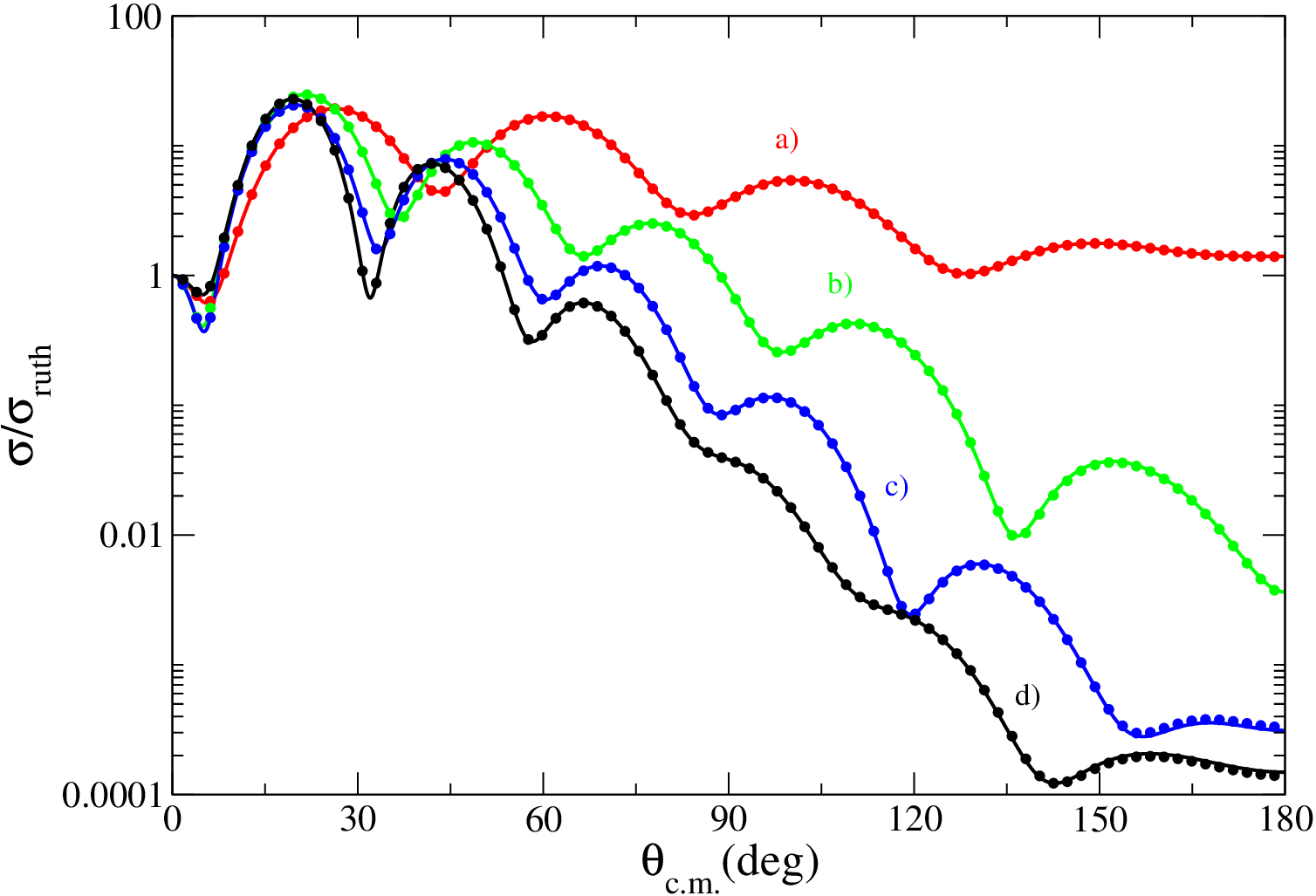}
}
\caption{(Color online) Differential cross section relative to Rutherford cross section for $p$-${\rm^{16}O}$ elastic scattering. Full lines are partial-wave calculations in configuration space and points are plane-wave calculations in momentum space. The lines a), b), c), and d) correspond to $E_{lab}=$50, 100, 150, and 200 MeV, respectively.}
\label{p16O_elR}
\end{figure}

\begin{figure}
\resizebox{0.45\textwidth}{!}{%
  \includegraphics{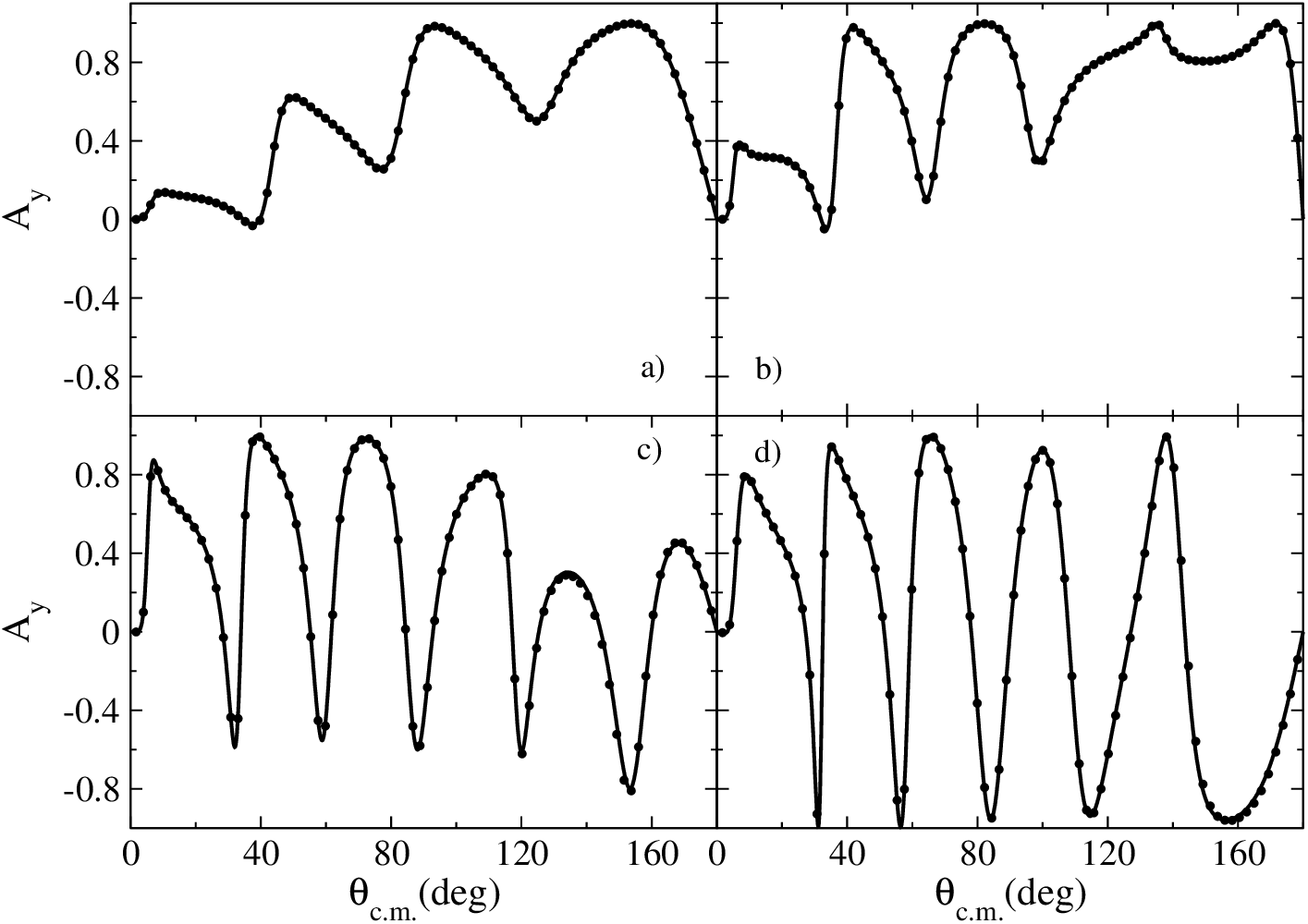}
}
\caption{Analyzing power for $p$-${\rm^{16}O}$ elastic scattering. Full lines are partial-wave calculations in configuration space and points are plane-wave calculations in momentum space. The lines a), b), c), and d) correspond to $E_{lab}=$50, 100, 150, and 200 MeV, respectively.}
\label{p16O_ap}
\end{figure}

Then, in Figs.~\ref{p10Be_elR} and \ref{p16O_elR}, we show the differential elastic cross section relative to the Rutherford cross section for these two reactions at the energies considered. The full lines represent configuration space partial-wave calculations, and the points show the plane-wave calculations in momentum space. Both calculations are in very good agreement at all the different energies. For $p$-${\rm^{16}O}$ scattering we also show in Fig.~\ref{p16O_ap} the analyzing power $A_y$ at the same energies as before. Again we obtain a very good agreement with the configuration space calculations. The perfect agreement  between the two calculations indicates that the method of screening and renormalization can be used accurately in high partial waves unlike the methods used in Refs.~\cite{Vi74,Cr90}.

\begin{figure}
\resizebox{0.45\textwidth}{!}{%
  \includegraphics{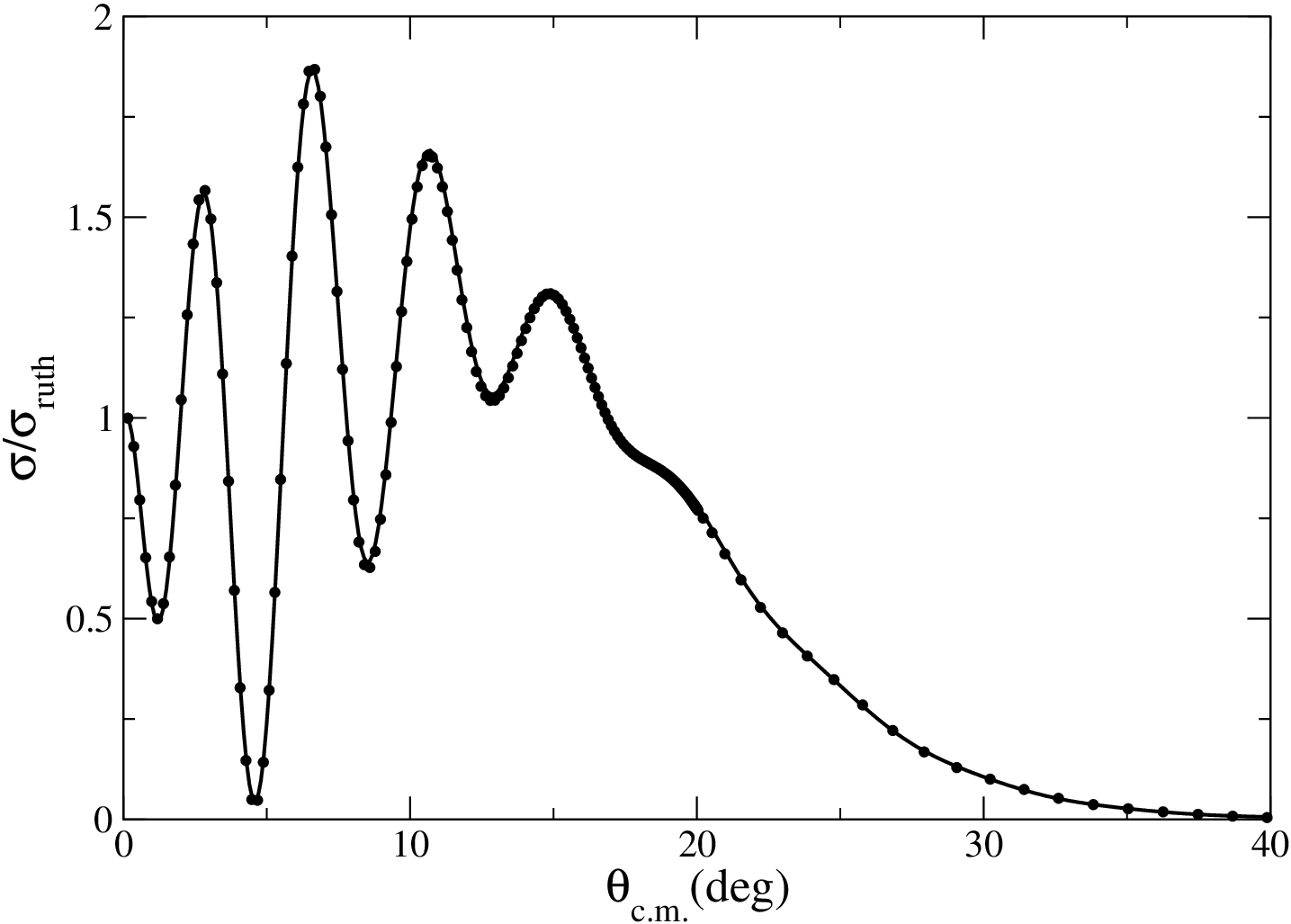}
}
\caption{Differential cross section relative to Rutherford cross section for  $\C$-${\rm^{10}Be}$@49.3MeV/u elastic scattering. Full line is partial-wave calculation in configuration space and points are plane-wave calculation in momentum space.}
\label{12C10Be_elR}
\end{figure}

Secondly we consider the reaction $\C$-${\rm^{10}Be}$. We study this reaction at 49.3 MeV per nucleon taking the optical potential used in Ref.~\cite{Alk97}. The screening radius is taken to be
$R=12$ fm and the number of mesh points used are $n_{\varphi}=$80, $n_q$=300, and $n_{\theta}$=240 divided in three regions in order to have more points where they are necessary. The CPU time needed for convergence on the same machine as above is now of about 1 hour using sixteen processors. The correspondent elastic scattering cross section relative to the Rutherford cross-section is shown in Fig.~\ref{12C10Be_elR}. The full line shows the configuration space partial-wave calculation,  and the points represent the plane-wave calculation in momentum space. Both results are again in very good agreement. However the convergence with the screening radius is slower than in the previous reactions and slower than using a partial-wave dependent renormalization factor as in Ref.~\cite{De05a}.

\section{Summary and Conclusions}

The two-body scattering has been studied by solving the Lippmann-Schwinger equation
in momentum space without partial-wave decomposition. The Coulomb and spin-orbit
interactions have been included. The screening and renormalization procedure has been used for including the Coulomb interaction.

The method has been applied to different reactions, $p$-${\rm^{10}Be}$, $p$-${\rm^{16}O}$ at 50, 100, 150, 200 MeV, and  $\C$-${\rm^{10}Be}$ at 49.3MeV/u. The results are in good agreement with those obtained through the solution of configuration space equations using the partial-wave decomposition. 

This work shows that this procedure is reliable for a two-body reaction and encourages its extension to three-body reactions where the partial wave decomposition becomes unstable at high energies or when two of the three particles are massive.

\begin{acknowledgments}
This work was supported  by the FCT under the grants POCTI/ISFL/2/275 and
POCTI/FIS/43421/2001. 
\end{acknowledgments}

\appendix

\section{Fourier transform} \label{ApendA}

In this Appendix we present the analytical expressions for the Fourier 
transform of the potentials used in this paper.

For a central potential $V(r)$, the Fourier transform is a scalar function
\begin{equation}
V(q',q,y)
=\frac{1}{(2\pi)^3}\int d^3rV(r)e^{-i({\bf q}-{\bf q}')\cdot{\bf r}},
\end{equation}
where $y$ is the cosine of the angle between ${\bf q}$ and ${\bf q'}$.
Integrating in the polar angles ($\theta,\varphi$), one gets
\begin{equation} \label{ftcp}
V(q',q,y) = \frac{1}{2\pi^2}\int_0^{\infty} dr r^2 V(r)\frac{\sin{(|{\bf q}-{\bf q'}|r)}}{|{\bf q}-{\bf q'}|r}.
\end{equation}
which involves a single $r$  integral.

\subsection{Screened Coulomb potential} \label{ApendA1}

If we define the screened Coulomb potential as in Eq.~(\ref{scpot}),
the Fourier transform can be performed analytically. 
Starting from Eq. (\ref{ftcp}), the integral is
\begin{equation}
\omega_R(q',q,y)=\frac{\alpha_eZ_pZ_t}{2\pi^2k^2}\lc 1-\frac{\pi^2\cos{(2kR)}\cos{(kR)}}{\pi^2-(2kR)^2} \rc,
\end{equation}
where  $k=|{\bf q}-{\bf q'}|$.
When $k\to0$ 
\begin{equation}
\omega_R(q',q,y)\to \frac{\alpha_eZ_pZ_t}{4\pi^2}(5-8/\pi^2)R^2.
\end{equation}
The renormalization phase $\phi_R$ is given in Ref.~{\cite{Tay75}, 
\begin{equation}
\phi_R= \kappa(q_0) \int_{1/2q_0}^{\infty} \omega_R(r)  dr,
\end{equation}
which for the $\omega_R(r)$ given in Eq. (\ref{scpot}) becomes 
\begin{eqnarray}
\lefteqn{\phi_R=\kappa(q_0) \left\{ \mbox{ln}(2q_0R)\right.}\\
&+&\left.\frac{1}{2}\lc\mbox{ln}(3)+Si\lp\frac{3\pi}{2}\rp-Si\lp\frac{\pi}{2}\rp\rc \right\},\nonumber
\end{eqnarray}
with
\begin{equation}
Si\lp\frac{3\pi}{2}\rp-Si\lp\frac{\pi}{2}\rp\approx0.23761058,
\end{equation}
and
\begin{equation}
\kappa(q_0) = \alpha_e Z_p Z_t \mu/q_0.
\end{equation}

The pure Coulomb amplitude $T_c$ needed in Eq.~(\ref{eq:T=Tc}) is given
as a function of the c.m. scattering angle $\theta_{c.m.}$  by
\begin{eqnarray}
T_c & = & \frac{\kappa(q_0)}{2(2\pi)^2 \mu  q_0 \sin^2\lp\frac12\theta_{c.m.}\rp}
\frac{\Gamma\lp1+i\kappa(q_0)\rp}{\Gamma\lp1-i\kappa(q_0)\rp} \nonumber \\
& \times & \exp{\lp -2i\kappa(q_0)\ln\lp\sin\lp\textstyle{\frac12} \theta_{c.m.}\rp\rp\rp}.
\end{eqnarray}

\subsection{Short-range Coulomb potential}

The Coulomb potential inside the nucleus is usually taken as the Coulomb potential for a uniformly charged sphere of radius $r_0$ whose difference from point Coulomb is
\begin{equation}
\omega_{CR}(r)=\alpha_eZ_pZ_t\lc\frac{1}{2r_0}\lp 3-\frac{r^2}{r_0^2}\rp-\frac{1}{r}\rc.
\end{equation}
Again starting form Eq.~(\ref{ftcp})
we get
\begin{eqnarray}
\lefteqn{\omega_{CR}(q',q,y)=-\frac{\alpha_eZ_pZ_t}{2\pi^2k^2}}\\
&\times&\frac{(kr_0)^3+3kr_0\cos{(kr_0)}-3\sin{(kr_0)}}{(kr_0)^3}.\nonumber
\end{eqnarray}
When $k\to0$
\begin{equation}
\omega_{CR}(q',q,y)\to -\frac{\alpha_eZ_pZ_t}{20\pi^2}r_0^2.
\end{equation}

\subsection{Woods-Saxon potential}
\label{ftws}

The Woods-Saxon potential, usually used for optical potentials, has the form
\begin{equation}
v_{ws}(r)=\frac{v_0}{1+e^{\frac{(r-r_0)}{a}}},
\end{equation}
leading to the Fourier transform 
\begin{equation}
v_{ws}(q',q,y)=\frac{v_0}{2\pi^2k}\int_0^{\infty} dr \frac{r\sin{(kr)}}{1+e^{\frac{(r-r_0)}{a}}}.
\end{equation}
Defining $x=r/a$, $b=r_0/a$, and $c=ka$ we have
\begin{equation}
v_{ws}(q',q,y)=\frac{v_0a^2}{2\pi^2k}\int_0^{\infty} dx \frac{x\sin{(cx)}}{1+e^{x-b}},
\end{equation}
which, except for the constant $v_0a^2/2\pi^2k$, may be calculated as the imaginary part of the integral $I_{(x)}$, along the $x$ axis,
\begin{equation}
I_{(x)}=\int_0^{\infty} dx \frac{xe^{icx}}{1+e^{x-b}}.
\end{equation}
In order to calculate this integral we consider the integral in the complex plane $I_{(z)}$ over the first quadrant
\begin{equation}
I_{(z)}=\oint dz \frac{ze^{icz}}{1+e^{z-n}}=I_{(x)}+I_{(y)},
\end{equation}
that equals the sum of the same integral over the positive axes $x$ and $y$ because the integration over the arc is zero. Applying the Cauchy's theorem we have
\begin{eqnarray}
I_{(z)}&=&\sum_n 2\pi i \mbox{Res}\lp\frac{ze^{icz}}{1+e^{z-b}}\rp_n\\
&=&2\pi i\sum_{n=0}^{\infty}\lp ze^{icz}e^{b-z}\rp_{z=b+i\pi(2n+1)}\nonumber\\
&=&2\pi e^{icb}\frac{e^{-c\pi}}{1-e^{-2c\pi}}\lc\pi\frac{1+e^{-2c\pi}}{1-e^{-2c\pi}}-ib\rc.\nonumber
\end{eqnarray}
Now we need to calculate $I_{(y)}$
\begin{equation}
I_{(y)}=\int_{i\infty}^0 dz \frac{ze^{icz}}{1+e^{z-b}}=\int_0^{\infty} dy \frac{ye^{-cy}}{1+e^{iy-b}}.
\end{equation}
Using the Taylor series for the function $f(y)=1/(1+y)$
\begin{eqnarray}
I_{(y)}&=&\sum_{n=0}^{\infty}\int_0^{\infty} dy~ye^{-cy}(-1)^ne^{(iy-b)n}\\
&=&\sum_{n=0}^{\infty}(-1)^ne^{-bn}\int_0^{\infty} dy~ye^{-y(c-in)}\nonumber\\
&=&\sum_{n=0}^{\infty}(-1)^ne^{-bn}\frac{c^2-n^2+2inc}{(n^2+c^2)^2}.\nonumber
\end{eqnarray}
This series converges very fast because it has a negative exponential increasing with $n$. 
Therefore our integral is
\begin{eqnarray}
\mbox{Im}[I_{(x)}]&=&\mbox{Im}[I_{(z)}-I_{(y)}]\\
&=&2\pi\frac{e^{-c\pi}}{1-e^{-2c\pi}}\lc\pi\sin{(cb)}\frac{1+e^{-2c\pi}}{1-e^{-2c\pi}}-b\cos{(cb)}\rc\nonumber\\
&-&2c\sum_{n=1}^{\infty}(-1)^ne^{-bn}\frac{n}{(n^2+c^2)^2} .\nonumber
\end{eqnarray}
When $k\to0$ 
\begin{eqnarray}
\frac{\mbox{Im}[I_{(x)}]}{k}&=&\frac{a}{3}(\pi^2b+b^3)\\
&-&2a\sum_{n=1}^{\infty}(-1)^ne^{-bn}\frac{1}{n^3} .\nonumber
\end{eqnarray}

\subsection{Derivative of Woods-Saxon potential}

The derivative of the Woods-Saxon potential is usually defined as
\begin{equation}
v^D_{ws}(r)=-4a\frac{dv_{ws}(r)}{dr}=4w_s\frac{e^{(r-r_s)/a}}{(1+e^{(r-r_s)/a})^2} .
\end{equation}
Now we follow the same procedure as in the former subsection but with a new function. Defining again $x=r/a$, $b=r_s/a$, and $c=ka$, we have
\begin{equation}
v^D_{ws}(q',q,y)=\frac{4v_sa^2}{2\pi^2k}\int_0^{\infty} dx \frac{e^{x-b}x\sin{(cx)}}{(1+e^{x-b})^2} ,
\end{equation}
from which we can define
\begin{equation}
I_{(z)}=\oint dz \frac{ze^{icz}e^{z-b}}{(1+e^{z-b})^2}=I_{(x)}+I_{(y)} ,
\end{equation}
as the integral over the first quadrant leading to 
\begin{eqnarray}
I_{(z)}&=&\sum_n 2\pi i \mbox{Res}\lp\frac{ze^{icz}e^{z-b}}{(1+e^{z-b})^2}\rp_n\\
&=&2\pi i\sum_{n=0}^{\infty}\lp (icz+1)e^{icz}e^{b-z}\rp_{z=b+i\pi(2n+1)}\nonumber\\
&=&2\pi i e^{icb}\frac{e^{-c\pi}}{(1-e^{-2c\pi})^2}\lc e^{-2c\pi}(1+c\pi)+(c\pi-1)\right.\nonumber\\
&-&\left.ibc(1-e^{-2c\pi})\rc.\nonumber
\end{eqnarray}
Now we need to calculate $I_{(y)}$
\begin{equation}
I_{(y)}=\int_{i\infty}^0 dz \frac{ze^{icz}e^{z-b}}{(1+e^{z-b})^2}=\int_0^{\infty} dy \frac{ye^{-cy}e^{iy-b}}{(1+e^{iy-b})^2} ,
\end{equation}
for which we use the Taylor series expansion of the function $f(y)=1/(1+y)^2$ to obtain
\begin{eqnarray}
I_{(y)}&=&\sum_{n=1}^{\infty}\int_0^{\infty} dy~ye^{-cy}(-1)^{n+1}e^{(iy-b)n}n\\
&=&\sum_{n=1}^{\infty}(-1)^{n+1}ne^{-bn}\int_0^{\infty} dy~ye^{-y(c-in)}\nonumber\\
&=&\sum_{n=1}^{\infty}(-1)^{n+1}e^{-bn}\frac{c^2-n^2+2inc}{(n^2+c^2)^2}. \nonumber
\end{eqnarray}
This series converges again very fast because it has a negative exponential increasing with $n$. 
Therefore our integral is given by
\begin{eqnarray}
\mbox{Im}[I_{(x)}]&=&\mbox{Im}[I_{(z)}-I_{(y)}]\\
&=&2\pi\frac{e^{-c\pi}}{(1-e^{-2c\pi})^2}\lc cb\sin{(cb)}(1-e^{-2c\pi})\nonumber\right.\\
&+&\left.\cos{(cb)\lp e^{-2c\pi}(1+c\pi)+(c\pi-1)\rp}\rc\nonumber\\
&-&2c\sum_{n=1}^{\infty}(-1)^{n+1}e^{-bn}\frac{n^2}{(n^2+c^2)^2} .\nonumber
\end{eqnarray}
When $k\to0$
\begin{eqnarray}
\frac{\mbox{Im}[I_{(x)}]}{k}&=&a(\frac{\pi^2}{3}+\pi b^2)\\
&-&2a\sum_{n=1}^{\infty}(-1)^{n+1}e^{-bn}\frac{1}{n^2} .\nonumber
\end{eqnarray}

\subsection{Spin-orbit potential}

The spin-orbit potential, given by Eq.(\ref{sop}), is not central so Eq.~(\ref{ftcp}) is not valid.
In this case, the Fourier transform of the spin-orbit term is  
\begin{equation}
\la {\bf q}'|v_{so}|{\bf q}\ra=\frac{1}{(2\pi)^3}\int d^3rd^3r'e^{-i{\bf q}'\cdot{\bf r}'}V_{so}({\bf r})\delta({\bf r}'-{\bf r})e^{i{\bf q}\cdot{\bf r}}.
\end{equation}
Taking into account the form of $V_{so}({\bf r})$ given in Eq.~(\ref{sop}) and the definition of ${\bf l}={\bf r}\times{\bf p}={\bf r}\times(-i\nabla)$, we have
\begin{eqnarray}
\la {\bf q}'|v_{so}|{\bf q}\ra&=&\frac{1}{(2\pi)^3}\int d^3rd^3r'e^{-i{\bf q}'\cdot{\bf r}'}v_{so}(r)\\
&\times&\delta({\bf r}'-{\bf r})\lc{\bf r}\times(-i\nabla)\rc\cdot{\bf \sigma}e^{i{\bf q}\cdot{\bf r}}.\nonumber
\end{eqnarray}
Integrating in ${\bf r}'$ and using the property $({\bf a}\times {\bf b})\cdot {\bf c}={\bf a}\cdot({\bf b}\times {\bf c})$ we get 
\begin{equation}
\la {\bf q}'|v_{so}|{\bf q}\ra=\frac{1}{(2\pi)^3}\int d^3r~v_{so}(r){\bf r}\cdot\lc{\bf q}\times{\bf \sigma}\rc e^{i({\bf q}-{\bf q}')\cdot{\bf r}}.
\end{equation}
Since $v_{so}(r)$ has the form
\begin{equation}
v_{so}(r)=-\frac{1}{r}\frac{dv_{ws}(r)}{dr},
\end{equation} 
one may write
\begin{equation}
v_{so}(r){\bf r}=-\nabla{v_{ws}(r)},
\end{equation}
leading to
\begin{equation}
\la {\bf q}'|v_{so}|{\bf q}\ra=\frac{-1}{(2\pi)^3}\int d^3r\lp\nabla v_{ws}(r)\rp\cdot\lc{\bf q}\times{\bf \sigma}\rc e^{i({\bf q}-{\bf q}')\cdot{\bf r}}.
\end{equation}
Integrating by parts 
\begin{eqnarray}
\lefteqn{\la {\bf q'}|v_{so}|{\bf q}\ra}\\
&=&\frac{-1}{(2\pi)^3}\lc\int d^3r \nabla \cdot \lp v_{ws}(r)\lc{\bf q}\times{\bf \sigma}\rc e^{i({\bf q}-{\bf q}') \cdot{\bf r}}\rp\right.\nonumber\\
&-&\left.i({\bf q}-{\bf q}')\cdot\lc{\bf q}\times{\bf \sigma}\rc\int d^3r~v_{ws}(r)e^{i({\bf q}-{\bf q}')\cdot{\bf r}}\rc\nonumber,
\end{eqnarray}
and applying Gauss' theorem,
\begin{eqnarray}
\lefteqn{\int d^3r \nabla\lp v_{ws}(r)\lc{\bf q}\times{\bf \sigma}\rc e^{i({\bf q}-{\bf q}') \cdot{\bf r}}\rp}\\
&=&\oint_{S}v_{ws}(r)\lc{\bf q}\times{\bf \sigma}\rc\cdot dS=0,\nonumber
\end{eqnarray}
together with
\begin{equation}
({\bf q}-{\bf q}')\cdot\lc{\bf q}\times{\bf \sigma}\rc=-{\bf q}'\cdot\lc{\bf q}\times{\bf \sigma}\rc=-\lc{\bf q'}\times{\bf q}\rc\cdot{\bf \sigma},
\end{equation}
we finally have 
\begin{equation}
\la {\bf q}'|v_{so}|{\bf q}\ra=\frac{-i}{(2\pi)^3}\lc{\bf q'}\times{\bf q}\rc\cdot{\bf \sigma}\int d^3r~v_{ws}(r)e^{i({\bf q}-{\bf q}')\cdot{\bf r}}.
\label{ftso}
\end{equation}
where the integral is equals to the Woods-Saxon Fourier transform developed in Subsection~\ref{ftws}.

\section{Solution method} \label{ApendB}

The integral equations for the transition amplitude with central interaction
 Eq.~(\ref{lse-3}) and including spin-orbit interaction Eq.~(\ref{lses-3}) are solved using
the well known method of  Pad\'e summation~\cite{pade} with a choice of 
 an appropriate mesh for each variable in Eq.~(\ref{lse-3}) or Eq.~(\ref{lses-3}). 
For the momenta $q$ we take a Gauss-Chebyshev mesh converting the interval $\lc-1,1\rc$ into $\lc0,\infty\rp$ via
\begin{equation}
q=b\sqrt{\frac{1+u}{1-u}},
\end{equation}
where b is a scale used to extend or compress the mesh. The scale used in this work is 5 fm$^{-1}$.
For the cosines $x$ we take the Legendre mesh in the interval $\lc-1,1\rc$. Sometimes, depending on the problem, it is more convenient to divide the $x$-interval in regions and define a Legendre mesh in each one with different number of points. This procedure allows to increase the number of mesh points where they are more necessary.  

The integral in $\varphi$ in the Eq.~(\ref{ipot}) and~(\ref{ipots}) is also calculated with a Legendre mesh in the interval $\lc0,2\pi\rc$. In the case of a central potential the integration can be done in the interval $\lc0,\pi\rc$ and multiplying by two. This is very useful if the particles do not have spin.


\bibliographystyle{apsrev}
\bibliography{./2b}

\end{document}